\input epsfig.sty
\documentstyle[12pt]{article}
\begin{document}

\def\PRL#1{{\it Phys.~Rev.~Lett.~}{\bf #1}}
\def\PRD#1{{\it Phys.~Rev.~}{\bf D#1}}
\def\PR#1{{\it Phys.~Rev.~}{\bf #1}}
\def\NPB#1{{\it Nucl.~Phys.~}{\bf B#1}}
\def\NP#1{{\it Nucl.~Phys.~}{\bf #1}}
\def\PLB#1{{\it Phys.~Lett.~}{\bf B#1}}
\def\ARNPS#1{{\it Ann.~Rev.~Nucl.~Part.~Sci.~}{\bf #1}}
\def\ZPhysC#1{{\it Z.~Phys.~}{\bf C#1}}
\def\PRepC#1{{\it Phys.~Rep.~}{\bf C#1}}
\def\ProgTP#1{{\it Prog.~Th.~Phys.~}{\bf #1}}
\def\ModPL#1{{\it Mod.~Phys.~Lett.~}{\bf A#1}}

\def\beq{\begin{equation}}
\def\eeq{\end{equation}}
\def\beqa{\begin{eqnarray}}
\def\eeqa{\end{eqnarray}}
\def\sm{Standard Model~}
\def\ps{$SU(4)\times SU(2)_L\times SU(2)_R$}
\def\cf{{\it cf.} }
\def\ie{{\it i.e.} }
\def\Slash{\hskip -.6em/}
\def\ul{\underline}

\def\Slash#1{%
   \setbox0=\hbox{$#1$}\dimen0=\wd0\setbox1=\hbox{/}\dimen1=\wd1%
   \ifdim\dimen0>\dimen1\rlap{\hbox to \dimen0{\hfil/\hfil}}#1%
   \else\rlap{\hbox to \dimen1{\hfil$#1$\hfil}}/\fi}%

\rightline{EFI 95-04}
\bigskip
\rightline{February 1995}
\bigskip
\bigskip

{\centerline{\bf RADIATIVELY GENERATED FERMION MASSES}}
{\centerline {IN $\bf SU(4)\times
SU(2)_L \times SU(2)_R$}}

\bigskip
\centerline{\it Mihir P. Worah}
\smallskip
\centerline{Enrico Fermi Institute and Department of Physics}
\centerline{University of Chicago, Chicago, IL 60637}
\bigskip

\bigskip

\centerline{\bf Abstract}

We propose a model based on the gauge group $SU(4)\times SU(2)_L\times
SU(2)_R$ where the Dirac masses of all the known fermions are
generated as one-loop radiative corrections. We are able to generate
realistic quark and lepton masses and mixings without a large hierarchy of
Yukawa couplings or extra symmetries. The neutrino masses,
which are see-saw suppressed, lie in the mass range favored
to solve the solar neutrino problem. The importance of threshold
corrections to tree-level mass relations in certain non-supersymmetric
GUTs is demonstrated.

\newpage

{\bf I. Introduction}
\bigskip

Nature displays an obvious symmetry between leptons and quarks. To
state a few of the more prominent ones:
they both come in three families, the negatively charged members of
corresponding quark and lepton families are similar in mass, and
there is a large hierarchy between the masses of the different
families.  The Standard Model
can account for all of these features: we understand from gauge
anomaly cancellation that the number of quark and lepton families has
to match up, and the fermion masses and their  hierarchies are
accounted  for by a hierarchy
of Yukawa couplings of the quarks and leptons to the Standard Higgs.
However, the features we mention
seem to ask for a unified explanation rather than just a way to
account for them.

Another prominent feature of the fermion mass spectrum is that,
compared with the other fermions in the same family, the neutrinos
are practically
massless. Although in the minimal Standard Model
the neutrinos are strictly massless, the on-going solar and
atmospheric neutrino experiments provide growing evidence for
small, but non-zero neutrino mass \cite{SNeut,ANeut}.
Certain cosmoligical models also prefer a non-zero neutrino
mass \cite{Cosmo}.
A  physically appealing explanation for the lightness of the
neutrinos is the see-saw mechanism \cite{GRSY}, which requires the
neutrinos to be Majorana particles. Thus the neutrinos differ in a
fundamental way from the rest of the fermions, all of which carry
conserved electromagnetic charge, and have to be Dirac particles.

Although there has been a host of recent work on predictions
for the fermion
masses based on unified theories \cite{DRH,NS,Ramond}, these models
usually rely on family symmetries to compute the Yukawa couplings as
functions of family charge and scale of family symmetry breaking.
We would like to build an extension of the \sm  that can
account for all the features we mentioned, while avoiding the extra
symmetries and scales of the models of Ref.~\cite{DRH,NS,Ramond}.
The model we propose unifies the
quarks and leptons, doesn't have a large hierarchy of Yukawa
couplings, and singles out the neutrinos as special. The
model has only two scales, the unification scale
$v_R$ and the electro-weak scale $v_L$, with $v_R\gg v_L$. Whatever
mechanism is responsible for this hierarchy in scales (manifested in our
present understanding as a finely-tuned Higgs potential) would also be
partially responsible for the hierarchies in fermion masses like
 $m_t\gg m_b,m_u$. It is not our
goal to predict the fermion masses and mixings,
but rather to show that one can
account for them in a simple model, without a proliferation of extra
symmetries and hierarchies.

A simple way to incorporate these ideas is to extend the \sm
gauge group $SU(3)\times SU(2)_L\times U(1)_Y$ \cite{GWS,GWP} to the
Pati-Salam group
$SU(4)\times SU(2)_L\times SU(2)_R$ \cite{PS,RABI1},
and have the fermion masses generated as one-loop radiative corrections
\cite{RABI2}.
The usual fermions transform under the gauge group as
$({\ul 4},{\ul2},{\ul1})$ or $({\ul 4},{\ul 1},{\ul 2})$, and we
introduce an extra sterile neutrino $s_0$ per generation. We choose
the Higgses to also transform as either $({\ul 4},{\ul2},{\ul1})$ or
$({\ul 4},{\ul 1},{\ul 2})$, which is a particularly simple and
attractive choice of Higgs representation.
When the neutral component of these Higgs
get vacuum expectation values the gauge group is broken in the
required way, but none of the fermions can get
a tree-level Dirac mass.
The sterile neutrino $s_0$ has a bare mass $m_0 \sim v_R$,
and it not only acts as a mass seed in generating one-loop
radiative masses for all the fermions, but also acts as the see-saw
partner of the left handed neutrino, suppressing its mass relative
to that of the other fermions.

This mechanism of mass generation demonstrates the impact of threshold
corrections to tree level mass relations in certain GUTs. We generate
fermion mass operators upon integrating out the heavy fields in the
theory. The coefficients of these operators are usually expected to be
small. However, as our model shows, they can be varied enough to
encompass the entire spectrum of fermion masses.

In Sec.~2 we briefly discuss generic features of models of radiative
mass generation, and how our model incorporates them. Sec.~3 discusses what we
call the ``minimal model'' and is the heart of the paper. In Sec.~4 we
extend the minimal model to be able to account for all the
observed fermion masses and mixings and discuss the results obtained,
 while Sec.~5  concludes.

\bigskip

{\bf II. About Radiative Masses}
\medskip

In any model of radiative fermion mass, the generic diagram that
generates a fermion mass term at one-loop looks like Fig.~(1){\footnote{For a
more detailed overview, and discussion of specific examples
see Refs.~\cite{Ibanez,NOP,BabuMa}.}}
\begin{center}
\epsfig{file=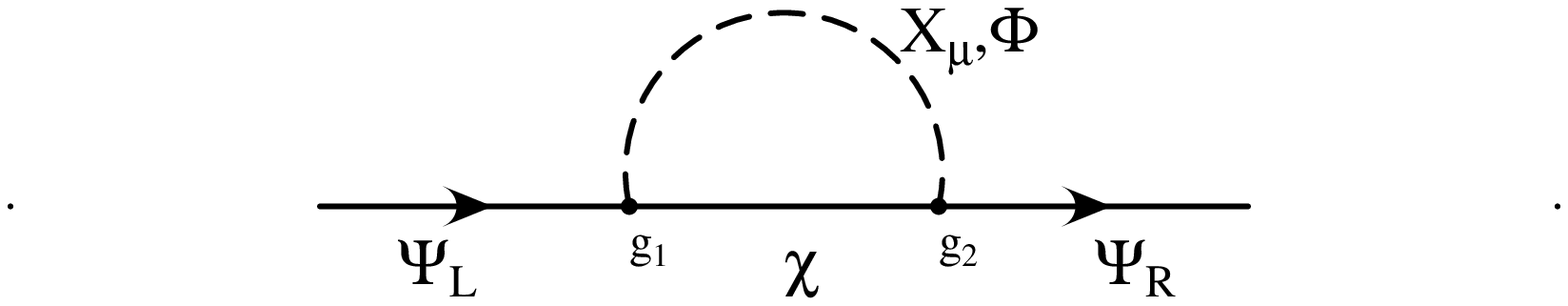,width=6in}
{\small {\bf Fig. 1.}  The general diagram for radiative fermion mass
generation.}
\end{center}
\vspace{0.5cm}
where the external lines correspond to the standard fermions
$\Psi_L,~\Psi_R$ (massless
at tree-level), the dashed line could be a scalar $\Phi$
or gauge boson $X_{\mu}$, and the solid
internal line is a (massive) fermion $\chi$. Since the representation content
of the theory is such that there can be no tree-level mass terms
connecting $\Psi_L$ to $\Psi_R$,
the sum of all such diagrams
gives a finite and calculable contribution to the external fermion
mass of order
\beq
m_{\psi}\sim \frac{g_1g_2}{(4 \pi)^2} v_L f(m_{\chi}/M_{X_{\mu},\Phi})
\label{est1}
\eeq
where $g_1$ and $g_2$ could be gauge or Yukawa couplings, $v_L$ is the
electro-weak scale, and $f(m_{\chi}/M_{X_{\mu},\Phi})$
is some kinematical function of the
masses on the internal line, with $f(m_{\chi}/M_{X_{\mu},\Phi})\leq 1$.

In order to give the \sm fermions mass we need:

 1) A chirality flip to turn the
incoming left handed fermion into an outgoing right handed fermion.
This is accomplished by having an odd number of mass insertions on the
fermion line. In our case the chirality flip is provided by the bare
mass of the  sterile neutrino $s_0$, and/or its mixings with the
standard neutrinos $\nu_L$ and $\nu_R$.

2) A change in weak isospin between left and right handed fermions.
In our model this is done either by $\nu_L$ and $\nu_R$ mixing with
$s_0$ on the
fermion line, or by Higgs of different weak isospin mixing on the scalar
line. These features are illustrated in Figs.~(2a,2b), and are
discussed in detail in the next section.

The original idea for radiative mass mechanism was to have some sort
of extra symmetries so that the third generation gets mass at tree-level, the
second at one-loop, and the third at two loop \cite{Weinberg}. These
models
are known as the $(1,\alpha,\alpha^2)$ models. In our model, we don't
impose any extra family symmetries, so all three generations get mass at
one-loop, and it would fit in more closely with  what are known as
$(\alpha,\alpha\epsilon,\alpha\epsilon^2)$ models.
(In the present case $\epsilon$ would roughly be some product of
ratios of Yukawa
couplings, Higgs mixing angles, and sterile neutrino masses).

One could in principle have supersymmetric models of radiative mass
generation. However, in this case the formula of Eq.~(\ref{est1})
gets modified ~\cite{ibanez2} to
\beq
m_{\psi}\sim \frac{g_1g_2}{(4 \pi)^2} \frac{M_{SUSY}^2}{v_R^2}v_L
              f(m_{\chi}/M_{X_{\mu},\Phi})
\label{est2}
\eeq
where $M_{SUSY}$ is the scale of supersymmetry breaking, and $v_R$ is
the high energy scale in the problem. To get around this suppression,
we need $M_{SUSY}\sim v_R$. Thus if $v_R$ is large, then $M_{SUSY}$
is forced to be large leaving the hierarchy problem unsolved, and if
$v_R$ is small, there is no hierarchy problem. In either case,
the primary motivation for supersymmetrizing a model is lost.

\bigskip
\newpage

{\bf III. The Minimal Model}

\medskip

In this section we would like to discuss a ``minimal'' version of the
proposed  model of radiative mass. This model contains only one
generation of fermions, and the minimal Higgs sector needed to achieve
the desired symmetry breaking. Although this minimal model is
not rich enough to describe the physical world,
 it is extremely simple to analyze, and differs from the full-blown
model in only the
trivial way that the full-blown model has additional replicas of the fermion
and scalar representations appearing here.

\medskip
{\bf A. Representation}
\medskip

As mentioned earlier, the gauge group we work with is $SU(4)\times
SU(2)_L\times SU(2)_R$.  We write the gauge fields in  the convenient
matrix notation, \ie we write $\hat V_{\mu} = \sum V_{\mu}^a\frac
{\lambda^a}{2} $ where $\lambda^a$ are generalized Gell-Mann matrices.
In this notation, we have the following gauge fields:
{\small
\beq
\hat W_{L\mu}  = \frac{1}{2}
                \left(\begin{array}{cc} W_{L\mu}^0& \sqrt 2 W_{L\mu}^+ \\
                \sqrt 2 W_{L\mu}^-& -W_{L\mu}^0 \end{array}\right)
\eeq
\beq
\hat W_{R\mu} = \frac{1}{2}
                \left(\begin{array}{cc} W_{R\mu}^0& \sqrt 2 W_{R\mu}^+ \\
                \sqrt 2 W_{R\mu}^-& -W_{R\mu}^0 \end{array}\right)
\eeq
%
\beq
\hat G_{\mu}  = \frac{1}{2}
\left(\begin{array}{cccc} G_{3\mu}+G_{8\mu}/\sqrt 3+B_{\mu}/\sqrt 6& \sqrt 2
G_{12\mu}^+& \sqrt 2 G_{13\mu}^+& \sqrt 2 X_{1\mu}^+\\
\sqrt 2 G_{12\mu}^-& -G_{3\mu}+G_{8\mu}/\sqrt 3+B_{\mu}/\sqrt 6& \sqrt 2
G_{23\mu}^+& \sqrt 2 X_{2\mu}^+\\
\sqrt 2 G_{13\mu}^-& \sqrt 2 G_{23\mu}^-& -2G_{8\mu}/\sqrt 3
+B_{\mu}/\sqrt 6& \sqrt 2 X_{3\mu}^+\\
\sqrt 2 X_{1\mu}^-& \sqrt 2 X_{2\mu}^-& \sqrt 2 X_{3\mu}^-&
-3 B_{\mu}/\sqrt 6 \end{array}\right).
\eeq
}
$G_{\mu}$ are the gluons, $B_{\mu}$ is the diagonal gauge boson that
couples to
$B-L$, and the $X_{\mu}$ are the lepto-quarks.

The fermions are in the usual representation of \ps~with the
addition of an extra sterile neutrino $s_0$,
\beqa
\Psi_{L i\alpha}& \sim   ({\ul 4},{\ul 2},{\ul 1}) \nonumber \\
\Psi_{R i\alpha}& \sim   ({\ul 4},{\ul 1},{\ul 2})\nonumber \\
s_0 & \sim  ({\ul 1},{\ul 1},{\ul 1}),
\eeqa
where $i=1,2$ is the $SU(2)_L$ or $SU(2)_R$ index, and $\alpha=1,2,3,4$ is the
$SU(4)$ index. Written out in matrix form, these fermion fields
would look like:
\beq
\Psi_{L,R} = \left(\begin{array}{cccc} u_1& u_2& u_3& \nu_e \\
                           d_1& d_2& d_3& e^- \end{array}\right)_{L,R}.
\eeq
We have one ``left handed'' Higgs $L$, and one ``right handed'' Higgs $R$
which transform as
\beqa
L_{ i\alpha}& \sim   ({\ul 4},{\ul 2},{\ul 1}) \nonumber \\
R_{ i\alpha}& \sim   ({\ul 4},{\ul 1},{\ul 2}).
\eeqa
and in matrix form,
\beq
L = \left(\begin{array}{cccc} L_{u1}& L_{u2}& L_{u3}& L_{\nu} \\
                           L_{d1}& L_{d2}& R_{d3}& L_{e} \end{array}\right)
\eeq
and
\beq
R = \left(\begin{array}{cccc} R_{u1}& R_{u2}& R_{u3}& R_{\nu} \\
                           R_{d1}& R_{d2}& R_{d3}& R_{e} \end{array}\right)
\eeq
If we embedded this model in $SO(10)$, the fermions $\Psi_L,~~\Psi_R$
would be
in a $\ul {16}$ in the usual way, the Higgs $L$ and $R$ would also be in a
$\ul {16}$, and there would be a single sterile neutrino
{\footnote {A model like this has been considered in Ref.~\cite {GeorgiN}.}}.

The symmetry breaking proceeds as
\beq
SU(4)\times SU(2)_L\times SU(2)_R\buildrel \langle R_{\nu}\rangle\over
\longrightarrow SU(3)\times SU(2)_L\times U(1)_Y\buildrel \langle
L_{\nu} \rangle\over \longrightarrow SU(3)\times U(1)_Q.
\eeq
So the gauge bosons of the broken groups
get tree-level masses by the usual Higgs mechanism, but the
representation content makes it impossible for the standard fermions
to get tree-level masses.

\medskip

{\bf B. Interactions}

\medskip

To proceed with the analysis of the model, we break up the interaction
Lagrangian into the following pieces: gauge-fermion, fermion-higgs,
higgs-higgs, higgs-gauge.

\medskip

{\underline {i. Gauge-Fermion.}}
\medskip

The gauge-fermion interaction is obtained from
\beq
i\bar\Psi_{L,R} \Slash D_{\mu}\Psi_{L,R}
\label{gferm1}
\eeq
where the covariant derivative is
\beq
D_{\mu}\Psi_{L,R}=\partial_{\mu}\Psi_{L,R}+
ig_4\hat G_{\mu}\Psi_{L,R}+ig_2\hat
W_{L,R\mu}\Psi_{L,R}
\eeq
This gives us the following  vertices in addition to the
$SU(3)_C$, $SU(2)_L$, and $SU(2)_R$ vertices,
\beqa
\cal L_{I} & = {\displaystyle{
-\frac{g_4}{2\sqrt 6}\bar u_L B^{\mu}\gamma_{\mu} u_L
-\frac{g_4}{2\sqrt 6}\bar d_L B^{\mu}\gamma_{\mu} d_L +
\frac{3g_4}{2\sqrt 6}\bar \nu_L B^{\mu}\gamma_{\mu} \nu_L +
\frac{3g_4}{2\sqrt 6}\bar e_L B^{\mu}\gamma_{\mu} e_L}} \nonumber \\
&
{\displaystyle{-\frac{g_4}{\sqrt 2}[\bar u_L X^{\mu}\gamma_{\mu} \nu_L
 + h.c.] -
\frac{g_4}{\sqrt 2}[\bar d_L X^{\mu}\gamma_{\mu} e_L + h.c.]}}
\nonumber \\
&
+{\displaystyle{L\leftrightarrow R}}.
\eeqa
\bigskip

{\underline{ii. Fermion-Higgs}
\medskip

The representation content of the model makes this sector extremely
simple:
\beq
{\cal L_{Y}} = -\kappa_L [\Psi_L^{i\alpha} L_{i\alpha} s_0 + h.c] -
              \kappa_R [\Psi_R^{i\alpha} R_{i\alpha} s_0^c + h.c].
\label{hferm1}
\eeq
Written out in components, this is
\beqa
\cal L_Y & = -\kappa_L[\bar u_L L_u s_0 + \bar d_L L_d s_0 +
              \bar \nu_L L_{\nu} s_0 + \bar e_L L_e s_0 +
h.c.]\nonumber \\
& -\kappa_R[\bar u_R R_u s_0^c + \bar d_R R_d s_0^c +
              \bar \nu_R R_{\nu} s_0^c + \bar e_R R_e s_0^c +
h.c.].
\label{hferm2}
\eeqa
When $L_{\nu}$ and $R_{\nu}$ get vacuum expectation values
$v_L/\sqrt 2$ and $v_R/\sqrt 2$  we get the
following tree-level neutrino mass matrix
\beq
M_{\nu} = \left(\begin{array}{ccc}0& 0& \kappa_L v_L/\sqrt 8 \\
                             0& 0& \kappa_R v_R/\sqrt 8 \\
                             \kappa_L v_L/\sqrt 8& \kappa_R v_R/\sqrt
				8 & m_0
                            \end{array}\right),
\label{nmass1}
\eeq
where $m_0\sim v_R$ is the bare mass of the sterile neutrino, and we've chosen
the basis ($\nu_L^c, \nu_R, s_0$). This matrix will be diagonalized by an
orthogonal matrix $\hat O$, with matrix elements $O_{ij}$,
 and will have eigenvalues $m_1=0,~m_2,~m_3\sim v_R$. So the left
handed neutrino is massless at tree-level. It will get a Dirac mass at
one loop like the rest of the fermions, but its physical mass
will be see-saw suppressed and much smaller than that of the other
fermions.

\medskip

{\underline{iii. Higgs-Higgs}.
\medskip
\beqa
V(L,R) &= -2\mu_L^2 L_{i\alpha}L^{i\alpha} +
\lambda_{L1}L_{i\alpha}L^{i\alpha} L_{j\beta}L^{j\beta} +
\lambda_{L2}L_{i\alpha}L^{j\alpha} L^{i\beta} L_{j\beta} +
\lambda_{L3}L_{i\alpha}L^{j\alpha} L^{i}_{\beta}L_{j}^{\beta} \nonumber \\
&
-2\mu_R^2 R_{i\alpha}R^{i\alpha} +
\lambda_{R1}R_{i\alpha}R^{i\alpha} R_{j\beta}R^{j\beta} +
\lambda_{R2}R_{i\alpha}R^{j\alpha} R^{i\beta} R_{j\beta} +
\lambda_{R3}R_{i\alpha}R^{j\alpha} R^{i}_{\beta}R_{j}^{\beta} \nonumber \\
& +
\lambda_{LR1}L_{i\alpha}L^{i\alpha} R_{j\beta}R^{j\beta} +
\lambda_{LR2}L_{i\alpha}R^{j\alpha} L^{i\beta} R_{j\beta} +
\lambda_{LR3}(L_{i\alpha}R^{j\alpha} L^{i}_{\beta}
R_{j}^{\beta}+h.c).
\label{minpot}
\eeqa
Here $L^{i\alpha}=(L_{i\alpha})^*$ and $L^{i}_{\alpha}=\epsilon^{ij}
L_{j\alpha} $.
In the limit of no L-R coupling, {\ie} $\lambda_{LR1}=\lambda_{LR2}=
\lambda_{LR3}=0$ we can generalize the arguments of Ref.~\cite{Li} to
show that if the following conditions are satisfied:
\beqa
\lambda_{L2} &< 0;~~\lambda_{L1}+\lambda_{L2} >
0;~~\lambda_{L3}>\lambda_{L2}~~or~~|\lambda_{L3}|>2\lambda_{L1}+\lambda_{L2}
\nonumber \\
\lambda_{R2} &< 0;~~\lambda_{R1}+\lambda_{R2} >
0;~~\lambda_{R3}>\lambda_{R2}~~ or~~
|\lambda_{R3}|>2\lambda_{R1}+\lambda_{R2}
\eeqa
the absolute minimum of the potential is at
\beqa
\langle L\rangle &=\left(\begin{array}{cccc} 0& 0& 0& v_L/\sqrt 2 \\
                           0& 0& 0& 0 \end{array}\right)\nonumber \\
\langle R\rangle &=\left(\begin{array}{cccc} 0& 0& 0& v_R/\sqrt 2 \\
                           0& 0& 0& 0 \end{array}\right)
\eeqa
This minimum will be stable to turning on the L-R couplings when
\beq
\lambda_{LR2} < 0;~~
\lambda_{LR1}+\lambda_{LR2} =  0;~~\lambda_{LR3}<\lambda_{LR2}/2
\label{minlrc}
\eeq
(Note the conditions on $\lambda_{LR}$ are sufficient, but not
necessary conditions).

The minima are then determined from the equatons
\beq
(\lambda_{L1}+\lambda_{L2})v_L^2+\frac{1}{2}(\lambda_{LR1}+
\lambda_{LR2})v_R^2=2\mu_L^2
\label{minconstr1}
\eeq
and
\beq
\frac{1}{2}(\lambda_{LR1}+\lambda_{LR2})v_L^2
+(\lambda_{R1}+\lambda_{R2})v_R^2=2\mu_R^2 .
\label{minconstr2}
\eeq
The parameters will have to be fine
tuned to generate the hierarchy $v_R\gg v_L$. There will be in general
only one light neutral Higgs, with the rest of the physical Higgses
having mass $\sim v_R$.

\medskip

{\underline{iv. Higgs-Gauge}}
\medskip

Not much detail is required here, except that given the symmetry breaking
scheme assumed, we get the following masses for the charged gauge bosons:
\beq
M_X^2
=\frac{g_S^2}{4}[v_R^2+v_L^2];~~M_{W_L}^2=\frac{g_L^2}
{4} v_L^2;~~M_{W_R}^2=\frac{g_R^2}{4} v_R^2~~~~~~~~~~~~
\label{cbmass}
\eeq
The neutral gauge bosons have the mass squared matrix
\beq
M_{0} = \frac{1}{8}\left(\begin{array}{ccc}
                   g_L^2v_L^2& 0& -(3/\sqrt 6)g_Lg_Sv_L^2 \\
                    0& g_R^2v_R^2& -(3/\sqrt 6)g_Rg_Sv_R^2 \\
               -(3/\sqrt 6)g_Lg_Sv_L^2& -(3/\sqrt 6)g_Rg_Sv_R^2&
                (3/2)g_S^2(v_R^2+v_L^2)
                            \end{array}\right)
\eeq
in the basis ($W_L^0,W_R^0,B^0$), with eigenvalues
\beq
M_{\gamma}^2=0;~M_Z^2=\frac{v_L^2}{4}[g_L^2+\frac{3g_R^2g_S^2}
{2g_R^2+3g_S^2}+{\cal O}(v_L^2/v_R^2)];
{}~M_{Z'}^2=\frac{v_R^2}{8}[2g_R^2+3g_S^2+{\cal O}(v_L^2/v_R^2)]
\label{nbmass}
\eeq
and eigenvectors
\beq
\left(\begin{array}{c}
\gamma \\ Z \\ Z' \end{array}\right)=
\left(\begin{array}{ccc}
s_W & s_W & \sqrt{c_{2W}} \\
c_W & -t_Ws_W & -t_W\sqrt{c_{2W}} \\
0 & -\sqrt{c_{2W}}/c_W & t_W \end{array}\right)
\left(\begin{array}{c}
W_L^0 \\ W_R^0 \\ B_0 \end{array}\right).
\eeq
For $v_L\ll v_R$,
the usual electro-weak relation $M_W^2=M_Z^2 c_W^2$ is still maintained
where we define
\beq
g_L^2=\frac{e^2}{s_W^2} \Rightarrow s_W^2=\frac{3 g_R^2 g_S^2}
                         {3g_R^2 g_S^2+2g_L^2 g_R^2+3g_L^2 g_S^2}
\eeq
(All of these tree-level relations are defined at the unification scale
$v_R=M_U$). For the rest of this paper we will assume that
$g_L(M_U)=g_R(M_U)=g_2(M_U)$, in which case we have the relation
\beq
s_W^2(M_U)=\frac{1}{2}-\frac{\alpha(M_U)}{3\alpha_S(M_U)}
\eeq
If we use as inputs at the electroweak scale $\alpha^{-1}=128.5$,
$s_W^2=0.23$, and $\alpha_S^{-1}=8.33$, then using the one-loop
$\beta$ functions with only gauge boson and fermion contributions,
this gives us $M_U=v_R\sim 10^{14}$
GeV, with $\alpha_S^{-1}(M_U)\sim 40,~\alpha_L^{-1}=\alpha_R^{-1} \sim
45$.

\bigskip

{\bf C. Fermion Masses}.
\medskip

{}From the interaction vertices we have written down we can see that
the processes that generate one-loop masses for the fermions are
leptoquark gauge boson exchange (Fig.~(2a)), which generates mass for
the up type quarks, Higgs exchange (Fig.~(2b)) which generates mass
for all the fermions, and neutral gauge boson exchange (Figs.~(3a,3b))
which generate Dirac and Majorana masses respectively for the neutrinos.

\newpage

\begin{center}
\epsfig{file=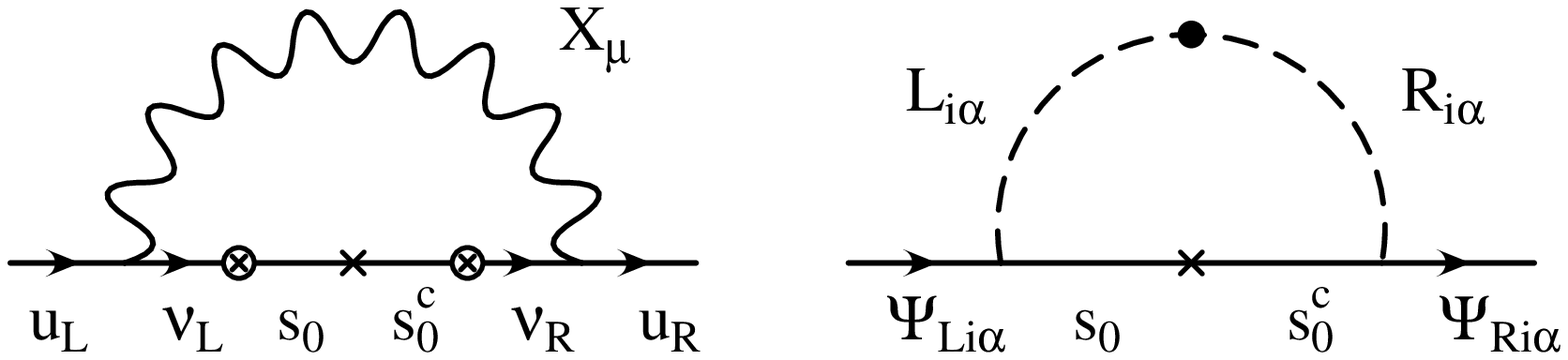,width=6in}
{\small {\bf Fig. 2a.}  Gauge boson exchange}.
\hskip 1.0in
{\small {\bf Fig. 2b.} Higgs exchange.}
\newline
{\small These diagrams are drawn in the interaction basis. $\times$
indicates a fermion mass insertion, $\circ$ indicates a change in left
or right weak isospin, and $\bullet$ indicates a change in both left
and right weak isospin.}
\end{center}
\vspace{0.5cm}

\begin{center}
\epsfig{file=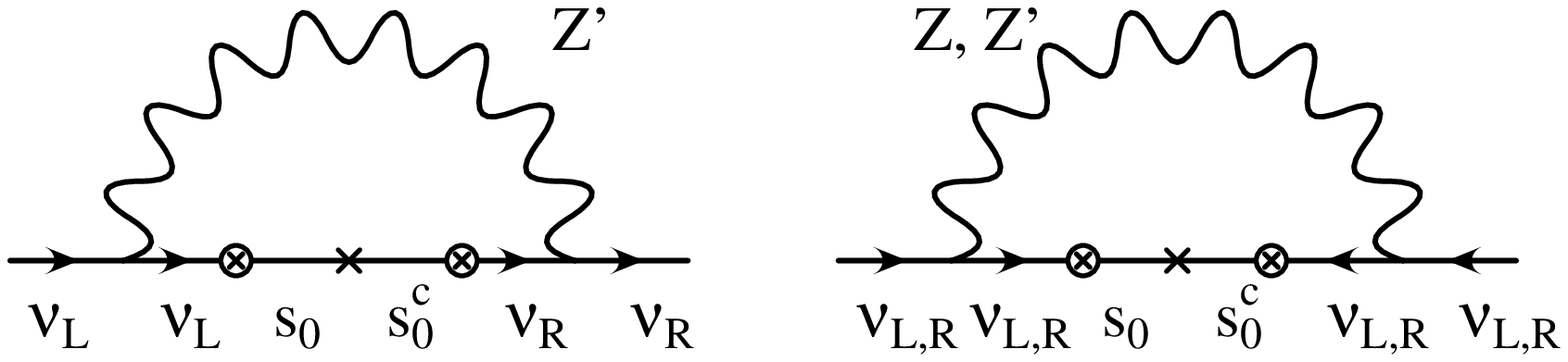,width=6in}
{\small {\bf Fig. 3a} Neutrino Dirac mass.}
\hskip 1.0in
{\small {\bf Fig. 3b} Neutrino Majorana mass.}
\end{center}
\vspace{0.5cm}

We see right away that this model gives us a way to differentiate
up type masses from down type masses.
The up type fermions get their masses from both gauge boson exchange and
Higgs exchange, whereas the down types get theirs only from Higgs exchange.

The results for the two types of diagrams calculated in
't Hooft-Feynman gauge are
\beq
m_{gauge}(M_U)=
2\frac{g_1g_2}{(4\pi)^2}\sum_{i=1}^3 O_{xi}O_{yi} m_i[\frac{m_i^2}{m_i^2-M_G^2}
                                  \ln(\frac{m_i^2}{M_G^2})]
\label{mgauge1}
\eeq
\beq
m_{Hj\alpha}(M_U)=\frac{\kappa_L\kappa_R}{(4\pi)^2}s_{j\alpha}
c_{j\alpha}
\sum_{i=1}^3 O_{3i}^2 m_i[
\frac{M_{j\alpha 1}^2}{m_i^2-M_{j\alpha 1}^2}
\ln(\frac{m_i^2}{M_{j\alpha 1}^2})-
\frac{M_{j\alpha 2}^2}{m_i^2-M_{j\alpha  2}^2}
\ln(\frac{m_i^2}{M_{j\alpha 2}^2})]
\label{mhiggs1}
\eeq
%
%
where $m_i$ are the masses of the physical neutrinos, and $O_{ij}$ are
elements of the orthogonal matrix that diagonalizes the neutrino mass
matrix~(\ref{nmass1}).  $M_G$ is the
mass of the appropriate
physical gauge boson, and $g_1,~g_2$ are its couplings to
the fermions. As an example, for the quarks, $g_1=g_2=g_s$, and $M_G=M_X$.
The subscripts x,y on the matrix $O$ depend
on the basis $(\nu_L^c,\nu_R,s_0)$ of Eq.~(\ref{nmass1}). For Dirac
masses x=1, y=2; left-handed Majorana mass x=1, y=1; right-handed
Majorana mass x=2, y=2.
 $M_{j\alpha 1,2}$ are the eigenvalues of the Higgs mass matrix
{\footnote{Unless some of the Higgs
are Nambu-Goldstone bosons, in which case one uses the mass of the
corresponding gauge boson.}}
in the basis $(L_{j\alpha}, R_{j\alpha})$,
and $s_{j\alpha}$ and $c_{j\alpha}$ are the sine and cosine of the
mixing angles between $L_{j\alpha}$ and $R_{j\alpha}$. Explicit
examples of Higgs mass matrices are presented in the
appendix (Eqs.~(\ref{mulr}-\ref{lamtdmass})). We would like to emphasize that
Eqs.~(\ref{mgauge1},\ref{mhiggs1})
are valid at the scale $M_U$. In order to compare
with the known fermion masses we need to use the renormalization
group to run the masses down to the appropriate scale.
Eqs.~(\ref{mgauge1},\ref{mhiggs1}) are the basis of all mass
calculations in this
paper, and we would like to first discuss some general features, then
illustrate their use with some examples.

The product $O_{1i}O_{2i}$, valid for Dirac masses,
in Eq.~(\ref{mgauge1}) represents mixing between
$\nu_L$ and $\nu_R$. One can tell by inspection of the mass matrix
Eq.~(\ref{nmass1}) that it is  of order $v_L/v_R$, while the masses $m_i$
and $M_X$ are of order $v_R$. Increasing the mass $m_i$ while holding
$v_L$ fixed simply
decreases the product $O_{1i}O_{2i}$. This is just a manifestation of the
decoupling theorem: one cannot generate arbitrarily large one-loop masses
$m_{gauge}$ by
increasing $m_i$ as a naive inspection of Eq.~(\ref{mgauge1}) seems to show.
In fact in order to maximize $m_{gauge}$ the masses $m_i$
are constrained to lie close to the unification scale $M_U$. A
numerical study shows that the maximum $m_{gauge}(M_U)$
can be is $\sim 10$ GeV.

In order to ensure the reliability of perturbative calculations, we
impose the following constraints on the Yukawa and Higgs couplings:
\beq
\frac{\kappa^2}{4\pi}<1\Rightarrow \kappa < 3.5;
{}~~~M_h^2<4v_R^2 \Rightarrow\lambda_i<4.
\label{condition}
\eeq
The Higgs mixing angle $s_{j\alpha}$ which represents mixing
between $L_{j\alpha}$ and $R_{j\alpha}$ is generically of order
$v_L/v_R$. One can in general fine tune the Higgs potential to
increase the mixing angle. However, a result of this fine tuning is
that the difference $(M_{j\alpha1}-M_{j\alpha2})\rightarrow 0$,
and the two terms in Eq.~(\ref{mhiggs1})
interfere destructively (this is illustrated in the appendix).
 Thus there is an upper bound to $m_{Higgs}(M_U)$
which is also around $10$ GeV.

One can also see that the up type quarks get contributions from both
gauge boson and Higgs exchange, while the down type quarks get a
contribution only from Higgs exchange. This feature will be important
in the full-blown model, as interference between different diagrams
will either enhance or suppress the up type quark masses compared with
the masses of the down type quarks.

Let us now consider an example:

If we choose as inputs
\beq
\kappa_L=\kappa_R=0.6;~~~m_0=0.4v_R;
\eeq
\beq
\begin{array}{ccc}
M_{u1}^2=0.05~v_R^2& M_{u2}^2=M_X^2& s_u=\epsilon \\
M_{d1}^2=0.05~v_R^2& M_{d2}^2=0.025~v_R^2& s_d=6~\epsilon
\end{array}
\label{mincs}
\eeq
where $\epsilon=v_L/v_R$, we get
\beq
m_c(100~GeV)=900~ MeV;~~~m_s(100~GeV)=100~ MeV.
\eeq
If we choose
\beq
\kappa_L=\kappa_R=0.04;~~~m_0=0.04v_R;
\eeq
\beq
\begin{array}{ccc}
M_{u1}^2=0.05~v_R^2& M_{u2}^2=M_X^2& s_u=\epsilon \\
M_{d1}^2=0.05~v_R^2& M_{d2}^2=0.025~v_R^2& s_d=80~\epsilon
\end{array}
\label{minud}
\eeq
we get
\beq
m_u(100~GeV)=5~ MeV;~~~m_d(100~GeV)=6~ MeV.
\eeq
All the Higgs masses and mixings were evaluated from the Higgs
potential of Eq.~(\ref{minpot}),
incorporating the constraints of
Eqs.~(\ref{minlrc}-\ref{minconstr2}, \ref{condition}).
This example serves to illustrate how a large hierarchy between the
charm and up masses is generated by several smaller hierarchies
in coupling constants and masses, and also that it is possible to
generate a large hierarchy in the up sector and simultaneously a
smaller hierarchy in the down sector by tuning the Higgs parameters.
Of course we are not free to use
different Higgs couplings for the different generations as we have
here. We use it here for illustrative purposes, as when we enlarge
the Higgs sector, a similar effect does
occur. Relative sizes of different Higgs self couplings determine
the sign and magnitude of the Higgs mixing angles, which in turn
determine whether different diagrams interfere constructively or
destructively.

An obvious problem with this model is that even with two families, the
couplings are all diagonal, and we never get inter-family mixing.
Another problem with this minimal model is that,
 as the arguments about upper bounds on
the masses show, we cannot generate a realistic mass for the
top quark.  Finally, this model cannot generate any mass at all for
the charged leptons. This is because the Higgs $L_e$ and $R_e$
are exact Nambu-Goldstone bosons (they are eaten by the $W_L$ and
$W_R$) and do not mix at tree-level.

Thus we see that although the ``minimal'' model serves as an attractive
and educative example of fermion mass generation, it is not rich
enough to describe the physical world. However, if one simply extends
the Higgs sector to include
another pair of Higgs $\Lambda$ and $T$ that transform exactly like
$L$ and $R$, all these problems disappear, and one can in fact
generate realistic fermion masses and mixings.
 This is the subject of the next section.

\bigskip

{\bf IV. The Full-Blown Model}

\medskip

In this section we would like to extend the minimal scenario of the
previous section, and present a complete model of
fermion masses and mixing. In order to accomplish this we first
introduce another pair of Higgs $\Lambda\sim (\ul 4,\ul 2,\ul 1)$,
and $T \sim (\ul 4,\ul 1,\ul 2)$. This simple addition to the model
greatly complicates the Higgs potential, and we relegate a detailed
discussion to the appendix. We should point out, however, that in
order for this model to give the charged leptons mass, and still
accomplish the correct gauge symmetry breaking, we must have the
vacuum expectation values $\langle \Lambda \rangle=\langle T
\rangle=0$. Thus the low energy data selects out a particular region
in the space of possible Higgs couplings (this is in some sense
analogous to the results in supersymmetric models of fermion mass
where the large $\tan
\beta$ case seems to be preferred by the low energy data~\cite{DRH}).
 As a result of this
particular choice of vacuum expectation values, the
formulas~(\ref{cbmass},\ref{nbmass}) for the gauge
boson masses remain unchanged. This also precludes the
possibility of introducing CP violation in the Higgs sector.

Next we generalize the Yukawa coupling constants to
matrices of couplings. The Higgs-fermion interaction Eq.~(\ref{hferm2}) now
looks like
\beqa
{\cal L_{Y}} &= -\kappa_L^{ab} [\Psi_{La}^{j\alpha} L_{j\alpha}
		s_{0b}+ h.c.]
 	 -\kappa_R^{ab} [\Psi_{Rb}^{j\alpha} R_{j\alpha}
		s_{0a}^c + h.c.]\nonumber \\
	&	-\kappa_{\Lambda}^{ab} [\Psi_{La}^{j\alpha} \Lambda_{j\alpha}
		s_{0b}+ h.c.]
 	 -\kappa_{T}^{ab} [\Psi_{Rb}^{j\alpha} T_{j\alpha}
		s_{0a}^c + h.c.],
\label{yukfull}
\eeqa
where $a,~b$ are family indices. The neutrino mass
matrix, and the orthogonal matrix $\hat O$ that diagonalizes it, will
now be $9\times 9$ matrices. The basis we will use in our subsequent
discussion groups all the left handed neutrinos first, followed by the
right handed, and finally the sterile neutrinos \ie $(\nu_{Le}^c,
\nu_{L\mu}^c,\nu_{L\tau}^c,\nu_{Re},\nu_{R\mu},\nu_{R\tau},
s_{0e},s_{0\mu},s_{0\tau})$.

The gauge-fermion interaction will still be diagonal, with
Eq.~(\ref{gferm1}) generalized to include diagonal family indices.

Finally, we come to the formula for fermion masses in the model. Most
of the work has already been done in the previous section, and all we
need to do is generalize Eqs.~(\ref{mgauge1},\ref{mhiggs1})
to sum over different
diagrams, and account for inter-family mixing.

\beq
m_{gauge}^{ab}(M_U)=
2\frac{g_1g_2}{(4\pi)^2}\sum_{i=1}^9
O_{x,i}O_{y,i} m_i[\frac{m_i^2}{m_i^2-M_G^2}
                                  \ln(\frac{m_i^2}{M_G^2})].
\label{mgfull}
\eeq
Here x=a, y=b+3 for Dirac masses; x=a, y=b for left handed Majorana
masses; x=a+3, y=b+3 for right handed Majorana masses.
\beqa
m_{LR,j\alpha}^{ab}(M_U)&={\displaystyle{
\sum_{m,n=1}^3\frac{\kappa_L^{am}\kappa_R^{bn}}{(4\pi)^2}
s_{LRj\alpha} c_{LRj\alpha}\sum_{i=1}^9 O_{m+6,i}O_{n+6,i}
m_i}}\nonumber \\
& {\displaystyle{[\frac{M_{LRj\alpha 1}^2}{m_i^2-M_{LRj\alpha 1}^2}
\ln(\frac{m_i^2}{M_{LRj\alpha 1}^2})
-\frac{M_{LRj\alpha 2}^2}{m_i^2-M_{LRj\alpha 2}^2}
\ln(\frac{m_i^2}{M_{LRj\alpha 2}^2})] }}
\label{mlrfull}
\eeqa
%
%
\beqa
m_{\Lambda T,j\alpha}^{ab}(M_U)&={\displaystyle{
\sum_{m,n=1}^3\frac{\kappa_\Lambda^{am}\kappa_{T}^{bn}}{(4\pi)^2}
s_{\Lambda Tj\alpha} c_{\Lambda Tj\alpha}\sum_{i=1}^9 O_{m+6,i}O_{n+6,i}
m_i}}\nonumber \\
& {\displaystyle{[\frac{M_{\Lambda Tj\alpha 1}^2}
{m_i^2-M_{\Lambda Tj\alpha 1}^2}
\ln(\frac{m_i^2}{M_{\Lambda Tj\alpha 1}^2})
-\frac{M_{\Lambda Tj\alpha 2}^2}
{m_i^2-M_{\Lambda Tj\alpha 2}^2}
\ln(\frac{m_i^2}{M_{\Lambda Tj\alpha 2}^2})] }}
\label{mltfull}
\eeqa

We would like to illustrate with an explicit example, that the model
can indeed generate realistic fermion mass matrices via
Eqs.~(\ref{mgfull},\ref{mlrfull},\ref{mltfull}). Our constraints are
to not have a large hierarchy of coupling constants, to keep coupling
constants below a magnitude where we can trust perturbation theory,
and finally to work honestly from the Lagrangian of the model. This
last point essentially states that we have to keep in mind the
relation between the Higgs masses and mixing angles: they are not
independent. As we show in the appendix, the mixing angle can be made
large only at the expense of lowering the differences in the Higgs
masses. This is a fact that is often not explicitly accounted for
 in papers on radiative mass generation.

Let us take the following as inputs to the model (all defined at the
scale $v_R\sim 10^{14}$ GeV).

i) Yukawa couplings.
\beq
\begin{array}{cc}
\kappa_L=\left(\begin{array}{ccc}
                 0.04& 0.03& 0.06\\
                 0.06& 0.42& 0.24\\
		 0.06& 0.08& 3.5\end{array}\right);
&\kappa_R=\left(\begin{array}{ccc}
                 0.04& 0.03& 0.06\\
                 0.06& 0.42& 0.24\\
		 0.06& 0.08& 3.5\end{array}\right)\\
\kappa_{\Lambda}=\left(\begin{array}{ccc}
                 -0.2& -0.12& -0.2\\
                 -0.26& -1.2& 0.28\\
		 3.5& 3.5& 3.5\end{array}\right);
&\kappa_T=\left(\begin{array}{ccc}
                 0.2& 0.1& 0.2\\
                 -0.34& 0.8& 0.18\\
		 3.5& 3.5& 3.5\end{array}\right)
\end{array}
\label{yukin}
\eeq

ii). Sterile neutrino bare masses.
\beq
m_{s_e}=0.5~v_R;~~~m_{s_{\mu}}=1.0~v_R;~~~m_{s_{\tau}}=7.0~v_R.
\label{sterin}
\eeq

iii). Higgs vacuum expectation values, masses, and mixings.
\beq
v_L=290~ GeV;~~~v_R=10^{14}~ GeV;~~~\epsilon=\frac{v_L}{v_R}.
\label{vevin}
\eeq

\beq
\begin{array}{ccc}
M^2_{LRu1}=2.0 ~v_R^2& M^{2'}_{LRu2}=M_X^2& s_{LRu}=\epsilon \\
M^2_{\Lambda Tu1}=2.0 ~v_R^2& M^2_{\Lambda Tu2}=0.25 ~v_R^2&
s_{\Lambda Tu}=1.14\epsilon \\
M^2_{LRd1}=2.0 ~v_R^2& M^2_{LRd2}=0.5 ~v_R^2& s_{LRd}=-0.03\epsilon \\
M^2_{\Lambda Td1}=2.0 ~v_R^2& M^2_{\Lambda Td2}=0.40 ~v_R^2&
s_{\Lambda Td}=0.05\epsilon \\
M^{2'}_{LRe1}=M_{W_R}^2& M^{2'}_{LRe2}=M_{W_L}^2& s_{LRe}=0 \\
M^2_{\Lambda Te1}=4.0 ~v_R^2& M^2_{\Lambda Te2}=1.0 ~v_R^2&
s_{\Lambda Te}=0.08\epsilon \\
M^2_{LR\rho 1}=M_{Z'}^2& M^2_{LR\rho 2}=M_Z^2& s_{LR\rho}=0 \\
M^{2'}_{LR\eta 1}=M_{Z'}^2& M^{2'}_{LR\eta 2}=M_Z^2& s_{LR\eta}=0 \\
M^2_{\Lambda T\rho 1}=3.8 ~v_R^2& M^2_{\Lambda T\rho 2}=3.0 ~v_R^2&
s_{\Lambda T\rho}=0.3\epsilon \\
M^2_{\Lambda T\eta 1}=3.8 ~v_R^2& M^2_{\Lambda T\eta 2}=3.0 ~v_R^2&
s_{\Lambda T\eta}=-0.3\epsilon
\end{array}
\label{lamin}
\eeq
All of these Higgs parameters are derived from the Higgs potential
(Eq.~(\ref{fullpot})),
keeping in mind the extremization equations
(Eqs.~(\ref{extrem1}-\ref{nomix2})), and the constraints on
the sizes of the couplings (Eq.~(\ref{condition})).
The mass values that are primed correspond
to the Nambu-Goldstone bosons, and the masses of the corresponding
gauge bosons are given in Eqs.~(\ref{cbmass},\ref{nbmass}). These masses, and
the mixing angles for the Nambu-Goldstone bosons are set by
Eqs.~(\ref{extrem1}-\ref{nomix2}) and cannot be adjusted. Notice that
we have chosen $v_L(v_R)=290~GeV$ to account for its running.
Given this choice of inputs we get the following outputs at the
electroweak scale ($\sim 100$ GeV)
\beq
m_u=1~MeV;~~~m_c=1~GeV;~~~m_t=180~GeV.
\label{upqEW}
\eeq
\beq
m_d=2~MeV;~~~m_s=100~MeV;~~~m_b=4~GeV.
\label{dnqEW}
\eeq
The absolute values of the quark mixing matrix are
\beq
|V_{KM}|=\left(\begin{array}{ccc}
                 0.98& 0.2& 0.05\\
                 0.2& 0.98& 0.1\\
		 0.05& 0.1& 0.99\end{array}\right)
\label{Vkm}
\eeq
\beq
m_e=0.1~MeV;~~~m_{\mu}=60~MeV;~~~m_{\tau}=3~GeV.
\label{elm}
\eeq
\beq
m_{\nu_e}=1\times 10^{-4}~eV;~~~m_{\nu_{\mu}}=3\times 10^{-3}~eV;~~~
m_{\nu_{\tau}}=4\times 10^{-2}~eV.
\label{nem}
\eeq
The absolute values of the lepton mixing matrix are
\beq
|V_{\nu}|=\left(\begin{array}{ccc}
                 0.99& 0.1& 0.05\\
                 0.1& 0.98& 0.15\\
		 0.05& 0.15& 0.99\end{array}\right)
\label{numix}
\eeq
As an example, let us explore how the top-bottom mass hierarchy arises
in this model. The top quark gets an ${\cal O}(10)$ GeV contribution
from gauge boson exchange, ${\cal O}(10)$ GeV contributions from each of
the 3 diagrams involving $L-R$ Higgs exchange, and additional
${\cal O}(40)$ GeV contributions from the 3 diagrams with $\Lambda - T$
Higgs exchange for a total mass of ${\cal O}(160)$ GeV.
On the other hand, the bottom quark gets no contribution from gauge
boson exchange, a total ${\cal O}(5)$ GeV from $L-R$ Higgs exchange,
and an {\em opposing} ${\cal O}(10)$ GeV from $\Lambda - T$ exchange for
a total of ${\cal O}(5)$ GeV. The exact expressions for the Higgs masses
and mixings are in Eqs.~(\ref{mulr}-\ref{lamtdmass}) of the appendix.

We would like to discuss what we have accomplished.
We started with a well motivated extension of the Standard Model,
 the Pati-Salam
group $SU(4)\times SU(2)_L\times SU(2)_R$. We add one sterile neutrino
per generation to the \sm fermion
content. The Higgs sector is
extremely simple, all Higgs fields transforming as $({\ul4},{\ul
2},{\ul 1})$ or $({\ul 4},{\ul 1},{\ul 2})$. The magnitude of every coupling
constant in the theory lies within the range
\beq
0.03<g_i,~\kappa_i,~\lambda_i<4,
\label{hierarchy}
\eeq
which is a hierarchy of $\sim 100$ (actually it's only the Yukawa
couplings that saturate this hierarchy; the Higgs couplings are all
within a factor of 10 of each other).
Certain Higgs self couplings are fine-tuned to give the symmetry
breaking we want, while the rest are assigned values within the
range of Eq.~(\ref{hierarchy}). The fermions receive masses as
one-loop radiative corrections.

All fermions within a generation have {\it identical} Yukawa couplings
to the Higgs~(\ref{yukfull}), but the Higgs to which the
different fermions couple may have different masses and mixings
depending on the Higgs potential. Using this fact, as well as the fact
that the up type quarks get their mass from gauge boson exchange in
addition to Higgs exchange, this model can generate the large
hierarchy $m_u/m_t\sim  10^{-5}$ while simultaneously
generating the much smaller hierarchy $m_d/m_b\sim 10^{-3}$. The
up-down mass inversion ($m_u<m_d$) is also achieved. The elements
of the quark mixing matrix have the correct order of magnitude as well
as hierarchies. Although the magnitudes of $V_{ub}$ and $V_{cb}$ are
larger than the experiment numbers, we postpone a search for more
realistic values till we have incorporated CP violation into this
model.

The charged leptons have
realistic masses, and the neutrino masses and lepton mixing matrix
have interesting values. For $\nu_{e}~-~\nu_{\mu}$ we have $\delta m^2\sim
10^{-5}~eV^2$, which lies in the range favored by the
MSW~\cite{MSW} solution to the solar neutrino problem~\cite{SNeut}.
These neutrino masses are fairly robust
since most of the neutrino mass comes from gauge boson exchange (this
is a natural consequence of the fact that the masses and mixings of
the neutral Higgses are constrained by their role in the spontaneous
symmetry breaking), however the mixing angle is dominated by the
charged lepton mixing, and dependent on the Yukawa couplings.

We think it to be extremely non-trivial, that such a simple extension
of the \sm can account for all of these features.
However, since we restricted
ourselves to real masses and coupling constants, there is no CP
violation in this model. It is our hope that we can generalize this
model to include sources of CP violation that can not only account for
the observed low energy CP violation in the $K$ meson system, but also
the observed baryon asymmetry. We are currently pursuing this
possibility, as well as studying possible low energy signatures of
this model~ \cite{MW}.

Although at this stage our
 aim is not to precisely reproduce the known values for
all of the observables (Eqs.~(\ref{upqEW}-\ref{numix})),
we should point out our approximations in
obtaining them. Eqs.~(\ref{mgfull}-\ref{mltfull}) were evaluated
using the inputs of Eqs.~(\ref{yukin}-\ref{lamin}) to obtain mass
matrices at the scale $v_R$. The mass matrices were numerically
diagonalized, giving the mass eigenvalues and mixing matrices also at
the scale $v_R$. In our approximation, none of the mixing angles or
the lepton masses vary with scale.
Each individual quark mass eigenvalue was scaled
using the one-loop $\beta$ functions with only gauge boson
and fermion contributions. Effectively this corresponded to an
enhancement of $\sim 2.6$ in each quark mass between the scales $v_R
\sim 10^{14}$ GeV and $M_W \sim 100$ GeV.
We estimate these approximations to introduce errors in the quark
masses of order
$20\% - 30\%$, in the lepton masses to be $\sim 5\%-
10\%$, and in the mixing angles $\sim 5\%$.

In a sense one could think of this model as demonstrating the
importance of threshold corrections to tree-level mass relations in
certain non-supersymmetric GUTs. We have started with the tree level mass
relation $m_{lepton}=m_{down}=m_{up}=0$, and essentially generated the
entire fermion mass spectrum as a consequence of the matching
conditions when we
integrate out the heavy fields! One could also envision the model
we have presented as being an intermediate scale effective theory of an
$SO(10)$ GUT, with sterile neutrinos and Higgs in the ${\ul {16}}$, as in
Ref.~\cite{GeorgiN}. In this case any tree-level masses the fermions
obtain from, say, Higgs in the ${\ul {10}}$ could be drastically modified by
threshold corrections (this possibility was in fact suggested in
Ref.~\cite{Ibanez,GeorgiN}). The bad news
is that tree-level relations such as $m_{\tau}=m_b$ may not scale as
naively expected. The good news is that it may not be necessary to
introduce extra Higgs like
${\ul{126}}$ solely to modify tree-level relations such as $m_e=m_d$.
We are currently investigating how the model presented here fits into
the $SO(10)$ framework~\cite{MW}.

One final comment we would like to make concerns the scale $v_R$ where
$SU(4)\times SU(2)_L\times SU(2)_R$ breaks to $SU(3)\times
SU(2)_L\times U(1)_Y$. So long as
$v_R \ge 10$ TeV, the mass formulas~(\ref{mgfull}-\ref{mltfull}) are
essentially independent of $v_R$.
This means that if we were willing to give up the condition
$g_L(v_R) = g_R(v_R)$ we could bring the scale $v_R$ down to the
several TeV range, as allowed by experiments~\cite{limits}. However
the largest top quark mass we can generate at the scale $v_R$ is about
80 GeV, so we need a large hierarchy between $v_L$ and $v_R$ in order
for it to scale by about the factor of two needed. In addition,
since the physical
neutrino masses are suppressed by the see-saw mechanism, a low $v_R$
would imply that
the neutrino masses are close to their direct experimental
upper bounds~\cite{PDG}, and we would have to give up the oscillation
solution to the solar neutrino problem.
These last two observations suggest that $v_R$ is indeed large, and
$\ge 10^{12}$ GeV.

\bigskip

{\bf V. Conclusions.}
\medskip

We have presented a model for radiative fermion mass based on
the gauge group $SU(4)\times SU(2)_L\times SU(2)_R$, with the fermion
content of the \sm extended to include one sterile neutrino per
generation.
We are able to generate realistic fermion masses
and mixings without a large hierarchy of coupling constants or extra
family symmetries. The electron and muon
neutrino masses lie in a range compatible with the
MSW solution to the solar neutrino problem.
The model demonstrates the potential importance of
threshold corrections to tree-level mass relations
in certain GUTs. The
possibility of CP violation and baryogenesis in this model are being
investigated.

\bigskip
{\bf VI. Acknowledgements.}
\medskip

The author would like to acknowledge useful discussions with Aaron
Grant and Jon Rosner. In addition he would like to thank Aaron for
computer advice, and Jon for a careful reading of the manuscript.
This work was supported in part by the United States Department of
Energy under Grant No. DE-FG02-90ER40560.

\bigskip

{\bf Appendix}
\medskip

In this appendix we would like to discuss the Higgs potential for the
full-blown model of Sec.~4. In order to do so we must
write down the most general $SU(4)\times SU(2)_L\times
SU(2)_R$ invariant potential involving the fields
$L_{i\alpha},~\Lambda_{i\alpha}\sim ({\ul 4},{\ul2},{\ul1})$ and
$R_{i\alpha},~T_{i\alpha}\sim ({\ul 4},{\ul1},{\ul2})$. This
potential will just be a long and complicated generalization of
Eq.~(\ref{minpot}), but before we write it out explicitly, we first define the
following $2\times 2$ matrix fields

\beq
\begin{array}{cccc}
{X^i}_j=L^{i\alpha}L_{j\alpha}&
{Y^i}_j=\Lambda^{i\alpha}\Lambda_{j\alpha}&
{Z^i}_j=L^{i\alpha}\Lambda_{j\alpha}& {} \\
\widetilde {X^i}_j=L^{i}_{\alpha}L_{j}^{\alpha}&
\widetilde {Y^i}_j=\Lambda^{i}_{\alpha}\Lambda_{j}^{\alpha}&
\widetilde {Z^i}_j=L^{i}_{\alpha}\Lambda_{j}^{\alpha}& {}\\
{A^i}_j=R^{i\alpha}R_{j\alpha}&
{B^i}_j=T^{i\alpha}T_{j\alpha}&
{C^i}_j=R^{i\alpha}T_{j\alpha}& {} \\
\widetilde {A^i}_j=R^{i}_{\alpha}R_{j}^{\alpha}&
\widetilde {B^i}_j=T^{i}_{\alpha}T_{j}^{\alpha}&
\widetilde {C^i}_j=R^{i}_{\alpha}T_{j}^{\alpha}& {}\\
{H^i}_j=L^{i\alpha}R_{j\alpha}&
{I^i}_j=L^{i\alpha}T_{j\alpha}&
{J^i}_j=\Lambda^{i\alpha}T_{j\alpha}&
{K^i}_j=\Lambda^{i\alpha}R_{j\alpha} \\
\widetilde {H^i}_j=L^{i}_{\alpha}R_{j}^{\alpha}&
\widetilde {I^i}_j=L^{i}_{\alpha}T_{j}^{\alpha}&
\widetilde {J^i}_j=\Lambda^{i}_{\alpha}T_{j}^{\alpha}&
\widetilde {K^i}_j=\Lambda^{i}_{\alpha}R_{j}^{\alpha},
\end{array}
\label{newfields}
\eeq
and as we have mentioned before, $i$ is an $SU(2)$ index, $\alpha$ is
the $SU(4)$ index, $\Psi^{i\alpha}=(\Psi_{i\alpha})^*$ and
$\Psi^{i}_{\alpha}= \epsilon^{ij}\Psi_{j\alpha}$. As an example,
\beq
\widetilde {H^i}_j=\left(\begin{array}{cc}L_dR^d+L^eR^e &
-(L_dR^u+L_eR^{\nu}) \\
-(L_uR^d+L_{\nu}R^e) & L_uR^u+L_{\nu}R^{\nu} \end{array}\right)
\eeq
Using the fields defined in Eq.~(\ref{newfields}) we have
\beqa
V(L,R,\Lambda,T)&=
-2\mu_{X}^2{X^i}_i-2\mu_Y^2{Y^i}_i-\frac{1}{2}
\mu_Z^2[{Z^i}_i+h.c.]\nonumber \\
&+\lambda_{XX1}{X^i}_i{X^j}_j+\lambda_{XX2}
{X^i}_j{X_i}^j+\lambda_{XX3}{X^i}_j\widetilde{X_i}^j\nonumber \\
&+\lambda_{YY1}{Y^i}_i{Y^j}_j+\lambda_{YY2}
{Y^i}_j{Y_i}^j+\lambda_{YY3}{Y^i}_j\widetilde{Y_i}^j\nonumber \\
&+\lambda_{XY1}{X^i}_i{Y^j}_j+\lambda_{XY2}
{X^i}_j{Y_i}^j+\lambda_{XY3}{X^i}_j\widetilde{Y_i}^j\nonumber \\
&+\lambda_{XZ1}[{X^i}_i{Z^j}_j+h.c.]+\lambda_{XZ2}
[{X^i}_j{Z_i}^j+h.c.]+\lambda_{XZ3}
[{X^i}_j\widetilde{Z_i}^j+h.c.]\nonumber \\
&+\lambda_{YZ1}[{Y^i}_i{Z^j}_j+h.c.]+\lambda_{YZ2}
[{Y^i}_j{Z_i}^j+h.c.]+\lambda_{YZ3}
[{Y^i}_j\widetilde{Z_i}^j+h.c.]\nonumber \\
&+\lambda_{ZZ}{Z^i}_i{Z_j}^j
+\lambda_{ZZ1}[{Z^i}_i{Z^j}_j+h.c.]+\lambda_{ZZ2}
{Z^i}_j{Z_i}^j+\lambda_{ZZ3}
[{Z^i}_j\widetilde{Z_i}^j+h.c.]\nonumber \\
&-2\mu_{A}^2{A^i}_i-2\mu_B^2{B^i}_i-\frac{1}{2}
\mu_C^2[{C^i}_i+h.c.]\nonumber \\
&+\lambda_{AA1}{A^i}_i{A^j}_j+\lambda_{AA2}
{A^i}_j{A_i}^j+\lambda_{AA3}{A^i}_j\widetilde{A_i}^j\nonumber \\
&+\lambda_{BB1}{B^i}_i{B^j}_j+\lambda_{BB2}
{B^i}_j{B_i}^j+\lambda_{BB3}{B^i}_j\widetilde{B_i}^j\nonumber \\
&+\lambda_{AB1}{A^i}_i{B^j}_j+\lambda_{AB2}
{A^i}_j{B_i}^j+\lambda_{AB3}{A^i}_j\widetilde{B_i}^j\nonumber \\
&+\lambda_{AC1}[{A^i}_i{C^j}_j+h.c.]+\lambda_{AC2}
[{A^i}_j{C_i}^j+h.c.]+\lambda_{AC3}
[{A^i}_j\widetilde{C_i}^j+h.c.]\nonumber \\
&+\lambda_{BC1}[{B^i}_i{C^j}_j+h.c.]+\lambda_{BC2}
[{B^i}_j{C_i}^j+h.c.]+\lambda_{BC3}
[{B^i}_j\widetilde{C_i}^j+h.c.]\nonumber \\
&+\lambda_{CC}{C^i}_i{C_j}^j
+\lambda_{CC1}[{C^i}_i{C^j}_j+h.c.]+\lambda_{CC2}
{C^i}_j{C_i}^j+\lambda_{CC3}
[{C^i}_j\widetilde{C_i}^j+h.c.]\nonumber \\
&+\lambda_{HH1}{X^i}_i{A^j}_j+\lambda_{HH2}
{H^i}_j{H_i}^j+\lambda_{HH3}[{H^i}_j\widetilde{H_i}^j+h.c.]\nonumber \\
&+\lambda_{II1}{X^i}_i{B^j}_j+\lambda_{II2}
{I^i}_j{I_i}^j+\lambda_{II3}[{I^i}_j\widetilde{I_i}^j+h.c.]\nonumber \\
&+\lambda_{JJ1}{Y^i}_i{B^j}_j+\lambda_{JJ2}
{J^i}_j{J_i}^j+\lambda_{JJ3}[{J^i}_j\widetilde{J_i}^j+h.c.]\nonumber \\
&+\lambda_{KK1}{Y^i}_i{A^j}_j+\lambda_{KK2}
{K^i}_j{K_i}^j+\lambda_{KK3}[{K^i}_j\widetilde{K_i}^j+h.c.]\nonumber \\
&+\lambda_{HI1}[{X^i}_i{C^j}_j+h.c.]
+\lambda_{HI2}[{H^i}_j{I_i}^j+h.c.]
+\lambda_{HI3}[{H^i}_j\widetilde{I_i}^j+h.c.]\nonumber \\
&+\lambda_{JK1}[{Y^i}_i{C^j}_j+h.c.]
+\lambda_{JK2}[{J^i}_j{K_i}^j+h.c.]
+\lambda_{JK3}[{J^i}_j\widetilde{K_i}^j+h.c.]\nonumber \\
&+\lambda_{HK1}[{Z^i}_i{A^j}_j+h.c.]
+\lambda_{HK2}[{H^i}_j{K_i}^j+h.c.]
+\lambda_{HK3}[{H^i}_j\widetilde{K_i}^j+h.c.]\nonumber \\
&+\lambda_{IJ1}[{Z^i}_i{B^j}_j+h.c.]
+\lambda_{IJ2}[{I^i}_j{J_i}^j+h.c.]
+\lambda_{IJ3}[{I^i}_j\widetilde{J_i}^j+h.c.]\nonumber \\
&+\lambda_{HJ1}[{Z^i}_i{C^j}_j+h.c.]
+\lambda_{HJ2}[{H^i}_j{J_i}^j+h.c.]
+\lambda_{HJ3}[{H^i}_j\widetilde{J_i}^j+h.c.]\nonumber \\
&+\lambda_{IK1}[{Z^i}_i{C^j}_j+h.c.]
+\lambda_{IK2}[{I^i}_j{K_i}^j+h.c.]
+\lambda_{IK3}[{I^i}_j\widetilde{K_i}^j+h.c.]
\label{fullpot}
\eeqa
We now assume the vacuum expectation values
\beq
\begin{array}{cc}
\langle L\rangle=\left(\begin{array}{cccc}
                 0& 0& 0&{\displaystyle{\frac{v_Lc_L}{\sqrt 2}}}\\
                 0& 0& 0& 0\end{array}\right);
&\langle\Lambda\rangle=\left(\begin{array}{cccc}
                 0& 0& 0& {\displaystyle{\frac{v_Ls_Le^{i\phi_L}}{\sqrt 2}}}\\
                   0& 0& 0& 0\end{array}\right)\\
\langle R\rangle=\left(\begin{array}{cccc}
                0& 0& 0& {\displaystyle{\frac{v_Rc_Re^{i\delta}}{\sqrt 2}}}\\
                  0& 0& 0& 0\end{array}\right);
&\langle T\rangle=\left(\begin{array}{cccc}
        0& 0& 0& {\displaystyle{\frac{v_Rs_Re^{i(\delta+\phi_R)}}{\sqrt 2}}}\\
                  0& 0& 0& 0\end{array}\right)
\end{array}
\label{VEV}
\eeq
and look for extrema of the potential.
Positivity of the Higgs masses will ensure that this extremum is
at least a local minimum.

In order for the charged leptons to get a one-loop mass, we need $L_e,~
\Lambda_e$ to mix with $R_e,~T_e$. However, these are the Higgses that
are eaten by the gauge bosons $W_L,~W_R$, which can't mix at tree
level in this model. Thus the only way the charged leptons can get a
mass is if some of the Higgses don't get a vacuum expectation value.
The choice we make is $\langle\Lambda\rangle=\langle T\rangle=0$, \ie
$s_L=s_R=0$ in Eq.~(\ref{VEV}). This choice also ensures that all the
potentially CP
violating phases $\phi_L,~\phi_R,~\delta$ are unphysical, and we set
them equal to 0. We are then left with the following four equations
that we need to satisfy in order to get the desired pattern of
symmetry breaking:
\beq
(\lambda_{XX1}+\lambda_{XX2})v_L^2+\frac{1}{2}(\lambda_{HH1}+\lambda_{HH2})
v_R^2=2\mu_X^2
\label{extrem1}
\eeq
\beq
\frac{1}{2}(\lambda_{HH1}+\lambda_{HH2})v_L^2+(\lambda_{AA1}+\lambda_{AA2})v_R^2
=2\mu_A^2
\label{extrem2}
\eeq
\beq
(\lambda_{XZ1}+\lambda_{XZ2})v_L^2
+(\lambda_{HK1}+\lambda_{HK2})v_R^2=4\mu_Z^2
\label{nomix1}
\eeq
\beq
(\lambda_{HI1}+\lambda_{HI2})v_L^2
+(\lambda_{AC1}+\lambda_{AC2})v_R^2=4\mu_C^2
\label{nomix2}
\eeq
Eqs.~(\ref{extrem1},\ref{extrem2}) which are identical to
Eqs.~(\ref{minconstr1},\ref{minconstr2}) of
Sec.~3 determine the scales $v_L,v_R$. One choice of parameters
consistent with the `minimal fine-tuning hypothesis' \cite{mft1,mft2} is
$\mu_X\sim v_L$, $\mu_A\sim v_R$ and
$\lambda_{HH1}+\lambda_{HH2}=0$.
In which case we get
\beq
v_L^2 = \frac{2\mu_X^2}{\lambda_{XX1}+\lambda_{XX2}}
\eeq
\beq
v_R^2 = \frac{2\mu_A^2}{\lambda_{AA1}+\lambda_{AA2}},
\eeq
with $\lambda_{AA1}+\lambda_{AA2},~\lambda_{XX1}
+\lambda_{XX2}\sim 1$.
Eqs.~(\ref{nomix1},\ref{nomix2}) ensure that $\langle\Lambda\rangle=\langle
T\rangle=0$. This also ensures that $L_e,~R_e$ don't mix with
$\Lambda_e,~T_e$. There will in general be one light neutral Higgs
(corresponding to the \sm Higgs boson),
with the rest having masses $\sim v_R$.

In order to simplify the calculation, we will adjust
the Higgs parameters to ensure a similar condition for the other Higgs
bosons \ie no mixing between $L-R$ and $\Lambda-T$ sectors. The
$4\times 4$ Higgs mass matrices now break up into $2\times 2$ blocks.
Consider as an example the mass squared matrices for
$L_u,~R_u,~\Lambda_u,~T_u$:
\beq
M_{LR}^{u}=-\frac{\lambda_{HH2}}{2}v_R^2\left(\begin{array}{cc}
           1& -\epsilon\\-\epsilon & \epsilon^2\end{array}\right).
\label{mulr}
\eeq
The masses and mixing angles are then
\beq
M^2_{LRu1}=-\frac{\lambda_{HH2}}{2}v_R^2;~~M^2_{LRu2}=0;~~
s_{LRu}=\epsilon,
\label{LRmass}
\eeq
where $\epsilon=v_L/v_R$.
Since we are calculating in 'tHooft-Feynman gauge, we use the gauge
boson mass for the mass of the Nambu-Goldstone boson. Thus, when
calculating the contribution from Eq.~(\ref{mlrfull}) we use
\beq
M'^2_{LRu2}=\frac{g_S^2v_R^2}{4}
\eeq
Notice that we have no freedom to vary the mixing angle $s_{LRu}$.
The mass squared matrix for $\Lambda_u,~T_u$ is
\beq
M_{\Lambda T}^{u}=\frac{v_R^2}{2}\left(\begin{array}{cc}
           \lambda_{KK1}& \lambda_{HJ2}\epsilon\\
\lambda_{HJ2}\epsilon & \lambda_{AB1}+\lambda_{AB2}\end{array}\right),
\eeq
with eigenvalues and mixing angles
\beq
M^2_{\Lambda Tu1}=\lambda_{KK1}\frac{v_R^2}{2};~~
M^2_{\Lambda Tu2}=(\lambda_{AB1}+\lambda_{AB2})\frac{v_R^2}{2};~~
s_{\Lambda Tu}=\frac{\lambda_{HJ2}}{\lambda_{KK1}-(\lambda_{AB1}+
\lambda_{AB2})}\epsilon.
\label{LTmass}
\eeq
Similarly the mass squared matrices for $L_d,~R_d,~\Lambda_d,~T_d$
are:
\beq
M_{LR}^{d}=\frac{v_R^2}{2}\left(\begin{array}{cc}
      \lambda_{HH1}& 2\lambda_{HH3}\epsilon\\
      2\lambda_{HH3}\epsilon &
2(\lambda_{AA1}+\lambda_{AA3})\end{array}\right),
\label{mdlr}
\eeq
with
\beq
M^2_{LRd1}=\frac{\lambda_{HH1}}{2}v_R^2;
{}~~M^2_{LRd2}=(\lambda_{AA1}+\lambda_{AA3})v_R^2;
{}~~s_{LRd}=\frac{2\lambda_{HH3}}{\lambda_{HH1}-2(\lambda_{AA1}+\lambda_{AA3})}
\epsilon .
\label{LRdmass}
\eeq
and
\beq
M_{\Lambda T}^{d}=\frac{v_R^2}{2}\left(\begin{array}{cc}
      \lambda_{KK1}& \lambda_{HJ3}\epsilon\\
      \lambda_{HJ3}\epsilon & \lambda_{AB1}+\lambda_{AB3}\end{array}\right),
\label{mdlamt}
\eeq
giving
\beq
M^2_{\Lambda Td1}=\frac{\lambda_{KK1}}{2}v_R^2;
{}~~M^2_{\Lambda Td2}=(\lambda_{AB1}+\lambda_{AB3})\frac{v_R^2}{2};
{}~~s_{\Lambda Td}=
\frac{\lambda_{HJ3}}{\lambda_{KK1}-(\lambda_{AB1}+\lambda_{AB3})}
\epsilon .
\label{lamtdmass}
\eeq

These are the values we use to calculate the contribution to the up
and down type quark masses from Eq.~(\ref{mltfull}). It would appear from
Eq.~(\ref{LTmass}) that we
can make the mixing angle as large as we want by tuning
$\lambda_{KK1}-(\lambda_{AB1}+\lambda_{AB2})$ to be small.
However as the mass terms in the same equation show, this would make
the Higgs degenerate in mass, and make the two terms in
Eq.~(\ref{mltfull}) cancel. Thus there is an upper limit to the
masses we can get in this model. It is interesting to note that even if we
saturate the coupling constants in this model to be largest they can
be consistent with maintaining a perturbative theory
(Yukawa couplings $\kappa\sim 3.5$, Higgs couplings $\lambda\sim 4$),
 the largest top
quark mass we can get in this model is $m_t\sim 200$ GeV!

\bigskip


\end{document}